\documentclass[11pt,a4paper]{article}
\usepackage[T1]{fontenc}
\usepackage{lmodern}
\usepackage{amsmath,amssymb,amsthm}
\usepackage{geometry,booktabs,array,hyperref,needspace}
\hypersetup{colorlinks=true,linkcolor=blue,urlcolor=blue}
\newtheorem{theorem}{Theorem}
\newtheorem{lemma}{Lemma}

\title{A detailed verification of Neary's PCP(5) construction}
\author{Vesa Halava\\
Department of Mathematics and Statistics\\University of Turku, Finland\\ Email: \href{mailto:vesa.halava@utu.fi}{vesa.halava@utu.fi}   }
\date{\today}
\begin{document}
\maketitle
\begin{abstract}
	Neary proved that the Post correspondence problem remains undecidable
	for instances with five pairs of words. We give a detailed verification
	of his construction. We first verify the undecidability of the
	restricted binary tag-system halting problem used in the reduction.
	We then prove both directions of the reduction to the five-pair
	problem, using the longer closing pair proposed by Rote. In particular,
	we show that every solution of the constructed PCP instance yields a
	halting computation of the simulated tag system. A companion Lean development verifies that the constructed PCP instance has exactly five pairs and admits a solution if and only if the source machine halts.
\end{abstract}

\section{Introduction}

An instance of the Post correspondence problem \cite{Post1946}
consists of a finite list of pairs of words \((u_i,v_i)\) over a
common alphabet. A solution is a nonempty sequence \(i_1,\ldots,i_t\),
with repetitions allowed, such that
\[
u_{i_1}\cdots u_{i_t}=v_{i_1}\cdots v_{i_t}.
\]
We write \(\mathrm{PCP}(k)\) for the restriction to instances
containing exactly \(k\) pairs. This article concerns the undecidability
of \(\mathrm{PCP}(5)\), already over the binary alphabet \(\{0,1\}\).

Matiyasevich and S\'enizergues established undecidability for a
three-rule semi-Thue system and thereby obtained a \(\mathrm{PCP}(7)\)
construction \cite{MS2005}. They used Claus's reduction from the
word problem of \(n\)-rule semi-Thue systems to
\(\mathrm{PCP}(n+4)\) \cite{Claus}. Neary proved
\(\mathrm{PCP}(5)\) undecidable using binary tag systems
\cite{Neary2015}. The purpose of this article is to give a detailed
verification of Neary's construction.

Neary's construction uses an encoding principle similar to that of
Matiyasevich and S\'enizergues. The program of the simulated system
is encoded in a single production. In Neary's binary tag system,
the right-hand side \(Ub\) of the rule \(c\to Ub\) encodes the
entire cyclic tag program, while the other rule is \(b\to b\).
The track at which \(Ub\) is read selects the appropriate instruction.

We give a detailed proof of both directions of Neary's five-pair
reduction, including the converse that a solution to the constructed PCP
instance implies that the binary tag computation halts.
Section~\ref{sec:tag} establishes the properties of the restricted tag
systems used in the reduction. We modify the doubling pass in the
Neary--Woods cyclic-tag simulation by replacing the appendants at
indices \(45\) and \(46\) in the lower half of Table~3.2 with
\(\mu\mu\). With this modification, the pass doubles the number of
live counters even when a block of counters precedes the doubling
marker \(D\). We then identify the designated halt instruction, show
how the binary tag system detects its firing, and prove that every
designated computation that terminates ends in the one-letter word
\(b\). Section~\ref{sec:pcp-instance} constructs the five pairs and
shows how a halting computation yields a PCP solution.
Section~\ref{sec:pcp-converse} proves the converse. The construction uses
Rote's longer closing pair \cite[Section~2.8, footnote~3]{Rote2025}.

A companion Lean development formalizes the reduction from the source-machine halting problem to binary PCP with five listed pairs. The formalization uses a local modification of the cyclic-tag construction, described before Lemma 1. The source code and instructions for checking the proofs are available in~\cite{leanPCP5}.

\section{The restricted binary tag systems}
\label{sec:tag}

We index the positions of a word from zero. Thus, for a word
\(v=v_0v_1\cdots v_{n-1}\) of length \(n\), \(v_i\) denotes the
symbol at position \(i\), where \(0\leq i<n\).

A tag system is a triple \(T=(A,R,\beta)\), where \(A\) is a finite
alphabet, \(R:A\to A^*\) assigns a production word to each letter,
and \(\beta\geq1\) is the deletion number. A configuration is a word
over \(A\), called the dataword. If the dataword is
\(a_0\cdots a_{n-1}\) with \(n\geq\beta\), one step gives
\[
a_0\cdots a_{n-1}\longmapsto a_\beta\cdots a_{n-1}R(a_0).
\]
The system halts when the dataword has length less than \(\beta\).
It is binary when its alphabet has two letters.

We construct the family used in the PCP reduction by simulating
a Turing machine first with a cyclic tag system and then with a
binary tag system. The cyclic simulation supplies a distinguished
instruction that signals halting.
A Turing machine \(M\) is deterministic and has a single tape indexed
by \(\mathbb Z\). It has a finite state set, an initial state, a
designated halting state, and a finite tape alphabet with a blank
symbol. For every nonhalting state and scanned symbol, the transition
function specifies a new state, a symbol to write, and a move one cell
to the left or right. On input \(y\), the letters of \(y\) occupy
consecutive cells starting at cell \(0\), all other cells are blank,
and the head starts at cell \(0\) in the initial state. The machine
halts upon entering its halting state.

\subsection{The cyclic halt signal}

\paragraph{Clockwise Turing machines.}
A binary clockwise Turing machine is a tuple
\(R=(Q_R,\Sigma_R,\delta_R,q_{\mathrm{in}},q_{\mathrm h})\),
where \(Q_R\) is a finite state set, \(\Sigma_R=\{A,B\}\) is the tape
alphabet, and \(q_{\mathrm{in}},q_{\mathrm h}\in Q_R\) are the
initial and halting states. Its transition function is
\[
\delta_R:(Q_R\setminus\{q_{\mathrm h}\})\times\Sigma_R
\longrightarrow Q_R\times(\Sigma_R\cup\Sigma_R^2).
\]
Represent a configuration by \((q,au)\), where \(au\) is the
nonempty tape word in clockwise order, starting at the scanned
symbol \(a\). If \(\delta_R(q,a)=(q',v)\), the next configuration
is \((q',uv)\). This replaces the scanned cell by the one or two
letters of \(v\) and places the head immediately after the
replacement in clockwise order. The machine halts in \(q_{\mathrm h}\).
The construction of Neary and Woods gives a binary clockwise
simulator \(R_M\) of \(M\). For each input \(y\), let
\(\tau_y\in\{A,B\}^{+}\) denote the initial tape word that this
construction gives for \(R_M\), written in clockwise order
starting at the scanned cell. Thus \(R_M\) halts from
\((q_{\mathrm{in}},\tau_y)\) if and only if \(M\) halts on \(y\)
\cite[Section~2 and Lemma~1]{NearyWoods2006}.

\paragraph{Cyclic tag systems.}
A cyclic tag program is a nonempty list of binary words
\(\alpha_0,\ldots,$ $\alpha_{p-1}\). A configuration is a pair
\((m,w)\), where \(0\le m<p\) is the current instruction
index and \(w\in\{0,1\}^*\) is the queue, with its front at
the left. Put \(m'=(m+1)\bmod p\). For each binary word \(v\),
the step relation is
\[
\begin{aligned}
	(m,0v)&\longrightarrow (m',v),\\
	(m,1v)&\longrightarrow (m',v\alpha_m).
\end{aligned}
\]
Thus a step reads and deletes the leftmost bit. If that bit is
\(1\), it appends \(\alpha_m\) at the right end, using the index
before the step. An empty queue admits no step. Instruction
\(h\) \emph{fires} precisely on a step from \((h,1v)\).

We use a local modification of the doubling pass in the cyclic-tag construction. In the lower half of Table 3.2, the appendants at indices \(45\) and \(46\) are replaced by \(\mu\mu\). This handles the layout in which live counters occur before \(D\): the repaired pass returns twice the original number of live counters. The simulation below uses these modified appendants throughout.

\begin{lemma}\label{lem:cyclic-halt}
	Given an ordinary single-tape Turing machine \(M\) and an input
	word \(y\), one can effectively construct a cyclic tag
	program \(\alpha_0,\ldots,\alpha_{p-1}\), a nonempty initial queue
	\(w\) at index zero, and a distinguished index \(0<h<p\) with
	the following properties.
	\begin{enumerate}
		\item Instruction \(h\) fires if and only if \(M\) halts on \(y\).
		If \(M\) does not halt, the cyclic computation has infinitely
		many steps.
		\item If the queue is \(1r\) just before the first firing of \(h\),
		then \(h\) does not fire during the next \(|r|\) steps after that
		firing, which consume the bits of \(r\).
	\end{enumerate}
\end{lemma}

\begin{proof}
	First construct \(R_M\) and the initial tape word \(\tau_y\)
	as above. The cyclic tag program simulates this clockwise machine.
	We use the construction in the proof of
	\cite[Section~3.1, Lemma~2]{NearyWoods2006}. Put \(N=|Q_R|\),
	\(z=30N+61\), and \(p=2z\). Number the states so that
	\(q_{\mathrm h}=q_N\).
	
	The (initial) configuration is encoded to the tag-system in the following way.
	The ordinary state and tape codewords and the counter word are
	\[
	\begin{aligned}
		S_i &= 0^{30i+20}1\,0^{p-30i-21},\\
		T_A &= 01\,0^{p-2}, & T_B &= 001\,0^{p-3},\\
		\mu &= 1\,0^{z-1}.
	\end{aligned}
	\]
	Extend \(T\) to words by concatenating the tape codewords.
	For a nonempty tape word \(\tau\), let \(K(\tau)\) be the
	least power of two at least \(|\tau|\). For
	\(q_i=q_{\mathrm{in}}\), the initial cyclic configuration is
	\[
	(0,w),\qquad w=S_i T(\tau_y)\mu^{K(\tau_y)}.
	\]
	This is equation~(3) of Neary and Woods.
	
	From~\cite{NearyWoods2006} we get the following details: 
	For every \(0\le n<p\), use the following rule. At the reserved
	index \(h=30N+20\) set \(\alpha_h=S_N\), as in the halting
	case of Neary and Woods. For all other indices use the assignments
	in their Tables~1.1, 1.2, 2.1, 3.1 and 3.2, substituting
	the states and transitions of \(R_M\). Each row specifies the
	index of the appendant and the binary word to assign to it.
	In Table~3.1 choose the transition row with the write length
	specified by \(\delta_R\) for the indicated state and tape
	symbol. Assign the empty word to every index left unassigned.
	These rules give the entire program effectively.
	
Neary and Woods simulate a clockwise transition in three stages.
The first two use the counter to mark and isolate the scanned tape
symbol. The third uses the current state and the isolated symbol to
select the transition of \(R_M\), writes the replacement symbols and
the new state, and restores the ordinary tape codewords. If the tape
grows beyond the current counter value, Table~3.2 doubles that value.
Each simulated transition takes finitely many cyclic tag steps and
returns the instruction index to zero.

For example, starting from \((0,S_iT(au)\mu^{K(au)})\), a transition
\(\delta_R(q_i,a)=(q_k,v)\) gives the boundary configuration
\[
(0,T(v)S_kT(u)\mu^{K(uv)}).
\]
Reading the tape codewords in circular order from \(S_k\) gives
the tape word \(uv\), starting at the new scanned cell.

In the halting case of the simulation
\cite[Section~3.1, end of the proof of Lemma~2]{NearyWoods2006},
the ordinary codeword for \(q_N=q_{\mathrm h}\) is entered at
index zero. Its unique \(1\) therefore fires instruction \(h\),
since
\[
S_N=0^h1\,0^{p-h-1}=\alpha_h.
\]

To see that $h$ cannot fire earlier, consider a cyclic step that reads
a~$1$ during the simulation of a nonhalting transition. For each encoded
object, Tables~1.1--3.2 of Neary and Woods give both the marker index
on entry and the appendant index at which the object's $1$ is read.
Thus it suffices to check the appendant-index column. Among the
state-dependent indices below $z$, the only ones congruent to $20$
modulo~$30$ are of the form $30i+20$. These occur when an ordinary state word \(S_i\) is entered at index
zero. The fixed indices below \(z\) are at most \(42<h\), and every
index at least \(z\) exceeds \(h\). In particular, when Table~3.2
copies a new state word \(S_k\) during counter doubling, the
selected index is \(30k+60\). Thus \(h\) fires exactly when the
simulation reaches the halting state. If \(R_M\) runs forever,
its successive transitions give infinitely many cyclic tag steps.
	
At the transition boundary where the halt state is first represented,
the queue contains a single state word $S_N$. Any tape words preceding
$S_N$ are read at index zero and copied to the rear before $S_N$ is
entered. When its unique $1$ fires instruction~$h$, the unread suffix
$r$ therefore consists of the $p-h-1$ trailing zeros of $S_N$ followed
by complete tape and counter words. After the trailing zeros the index is zero. Each tape
	word has length \(p=2z\) and has its \(1\) at offset \(1\) or \(2\).
	Each counter word is \(10^{z-1}\) and has length \(z\).
	Consequently the object boundaries are entered at index \(0\)
	or \(z\), and the \(1\)'s in these objects occur only at indices
	in \(\{0,1,2,z,z+1,z+2\}\), all distinct from \(h\).
	The new copy of \(S_N\) is appended behind \(r\),
	so no bit of that copy is read during the next \(|r|\) steps.
	This proves the second property.
\end{proof}

\subsection{The restricted halting problem}

The following construction uses the track method from Lemma~9 of
\cite{Neary2015} and the length congruence used in Neary's Theorem~11.

\begin{lemma}\label{lem:neary9}
	Given a Turing machine \(M\) and input \(y\), one can effectively
	construct a deletion number \(\beta\ge2\) and a binary tag system with
	productions \(R(b)=b\), \(R(c)=Ub\), where \(U=u_0\cdots u_\ell\).
	On the designated input
	\[
	w_0=u_{\beta-1}u_\beta\cdots u_\ell b
	\]
	the system halts if and only if \(M\) halts on \(y\).
	For the constructed \(\beta\), one may prescribe \(|Ub|=s\beta\)
	for every sufficiently large \(s\equiv1\pmod{\beta-1}\).
\end{lemma}

\begin{proof}
	We use Neary's method of storing the cyclic tag program in the tracks
	of one production word. We give every ordinary output the same initial
	padding. This allows a single track formula even for an empty cyclic
	appendant.
	
	Take the cyclic program \(\alpha_0,\ldots,\alpha_{p-1}\), initial
	word \(w\), and halt instruction \(h\) from Lemma~\ref{lem:cyclic-halt}.
	
    Put \(\beta=10p\) and write \(P=Ub\) for the production word to
    be constructed, of length \(s\beta\). For \(0\le j<\beta\), define
    its \(j\)-th track by
    \[
    T_j=P_jP_{j+\beta}\cdots P_{j+(s-1)\beta}.
    \]
    These tracks partition the positions of \(P\). A prescription for
	every \(T_j\) therefore determines \(P\). Extend the productions to a
	morphism \(\varphi\) by \(\varphi(b)=b,\ \varphi(c)=P\).
	If a traversal first reads position \(j\) of a copy of \(P\), it
	reads \(T_j\) and appends \(\varphi(T_j)\). More generally, the first
	read position in a block is called its offset. On leaving a block of
	length \(d\) entered at offset \(z\), the offset into the next block is
	\(z-d\bmod\beta\), since consecutive read positions differ by \(\beta\).
	Offsets in what follows are least nonnegative residues.
	
	Let \(G=P\), and define three further blocks and their short encodings by
	\[
	\begin{array}{c|c|c}
		& \text{short encoding} & \text{block}\\ \hline
		\text{padding}& e=b^4cb^6 & E=b^4Pb^6\\
		0 & x_0=b^6cb^4 & X_0=b^6Pb^4\\
		1 & x_1=b^8cb^2 & X_1=b^8Pb^2 .
	\end{array}
	\]
	For a bit word \(v\), write \(x(v)\) and \(X(v)\) for the corresponding
	concatenations. Thus \(\varphi(e)=E\) and \(\varphi(x(v))=X(v)\).
	We use \(G\) and complete groups \(E^p\) as padding. The offset
	representing program index \(m\) is
	\[
	z_m=(1-10m)\bmod\beta .
	\]
	Reading \(G\) preserves the offset. Each of \(E,X_0,X_1\) has length
	\(s\beta+10\), so it takes \(z_m\) to \(z_{m+1\bmod p}\).
	In particular \(E^p\) preserves the incoming index.
	
	We now prescribe the tracks used by the ordinary blocks.
	Take \(a\in\{4,6,8\}\) and enter \(b^aPb^{10-a}\) at offset \(z_m\).Inside \(P\) the track is \(T_j\), where \(j=(z_m-a)\bmod\beta\).
	For \(m=0\), the offset is \(1\), so the traversal reads one leading
	\(b\) followed by \(T_j\). For \(m>0\), the offset is at least \(11\)
	and it reads exactly \(T_j\). Indeed the block contains \(s+1\)
	sampled positions when \(z_m=1\), and \(s\) when \(z_m\ge11\).
	In both cases the \(s\) samples inside \(P\) exhaust all reads after
	the possible leading \(b\).
	
	Let \(\delta_m=1\) when \(m=0\), and \(0\) otherwise. Take
	\(v=\varepsilon\) for \(E\) and \(X_0\), and \(v=\alpha_m\) for
	\(X_1\) with \(m\ne h\). Prescribe the word read across the whole block as
	\[
	D=e^p x(v)c^t,\qquad t=s+\delta_m-11p-11|v|.
	\]
	The short encodings have length \(11\), so \(D\) has length
	\(s+\delta_m\). It begins with \(b\). For \(m>0\) put \(T_j=D\);
	for \(m=0\) remove the first \(b\) of \(D\) to obtain \(T_j\).
	In either case the block appends precisely
	\[
	\varphi(D)=E^p X(v)G^t .
	\]
	Thus even an empty \(v\) produces only complete padding. For \(G\)
	itself put \(T_{z_m}=c^s\), giving the output \(G^s\).
	
	For \(X_1\) at index \(h\) prescribe \(T_{(z_h-8)\bmod\beta}=bc^{s-1}\).
	As \(h>0\), the whole block reads this track and appends
	\(H=bG^{s-1}\). Set all even tracks equal to \(b^s\). Finally put
	\[
	T_{\beta-1}=b^{\beta-2}x(w)c^k e^p,\qquad
	k=s-(\beta-2)-11|w|-11p .
	\]
	The residues modulo \(10\) of the tracks for \(G,E,X_0,X_1\) are
	\(1,7,5,3\), respectively. Within each class the \(p\) program indices
	give distinct tracks modulo \(10p\). The input track has residue \(9\);
	all the even tracks are separate. The exceptional track takes the place
	of the \(X_1\) track at \(h\). Fill remaining tracks with \(b^s\).
	These assignments are effective. Take any \(s\equiv1\pmod{\beta-1}\)
	large enough that \(s\ge2\beta+3\), \(k\ge0\), and every ordinary
	exponent \(t\) is nonnegative. The last letter of \(P\) is the last
	letter of \(T_{\beta-1}\), namely \(b\). Hence \(P=Ub\) as required.
	
	\smallskip
	
	We are ready to study the computation of the binary tag system defined. 
	Deleting the first \(\beta-1\) letters of \(P\) gives exactly \(w_0\).
	A traversal of this suffix reads \(T_{\beta-1}\) and appends
	\(b^{\beta-2}X(w)G^kE^p\).
	The suffix has length \((s-1)\beta+1\), so the last deletion extends
	\(\beta-1\) places into the output. It consumes the prefix \(b^{\beta-2}\)
	and one letter of \(X(w)\). The first encoded bit is therefore entered
	at \(1=z_0\).
	
	To justify this and all subsequent block traversals, consider even the
	case in which no further block is waiting behind the current one.
	The initial suffix and every ordinary block, at every entry offset
	used here, contain at least \(s\) sampled positions. Up to the last
	sample each deletion fits inside the block, since another sample lies
	\(\beta\) positions ahead. By the last sample at least \(s-1\) letters
	have already been appended. They supply the at most \(\beta-1\)
	additional letters required by that deletion. This also applies to
	each copy of \(G\) traversed inside \(H\).
	
	Before the exceptional track is used, consider a boundary between
	successive traversals of units \(X_0,X_1,G,E^p\), treating each
	complete group \(E^p\) as one unit. The preceding deletion may
	already have consumed a prefix of the first pending unit. Its
	entry offset records the length of that prefix.
	
	The simulation invariant is that the pending units, with each
	\(G\) and \(E^p\) omitted and each \(X_i\) replaced by \(i\), give
	the current cyclic queue. If the current instruction is \(m\),
	the entry offset at the boundary is \(z_m\). Initialization
	establishes the invariant with queue \(w\) and offset \(z_0=1\).
	Traversing \(G\) or \(E^p\) preserves the offset and appends only
	padding. Traversing \(X_i\) advances the offset to
	\(z_{(m+1)\bmod p}\) and appends, apart from padding, the encoding
	of \(\alpha_m\) if \(i=1\) and of the empty word if \(i=0\),
	provided \((i,m)\ne(1,h)\). This proves the invariant by induction
	over the traversed units.
	
	Every pending encoded bit is reached after finitely many
	traversals. Only finitely many units precede it, and their outputs
	are appended behind it. Thus an infinite cyclic computation
	gives infinitely many legal binary tag steps.
	
	When instruction \(h\) first fires, the exceptional track appends
	\(H\). The encoded bits already ahead of \(H\) are precisely
	the bits of the suffix \(r\) in Lemma~\ref{lem:cyclic-halt}.
	Their instruction indices and actions are unaffected by the
	replacement of the cyclic output at the rear by \(H\).
	Lemma~\ref{lem:cyclic-halt} ensures that none of these bits fires
	\(h\). Only finitely many units precede \(H\), and their outputs
	are appended behind it, so \(H\) is reached after finitely many
	traversals. All the units then waiting behind \(H\) are ordinary
	ones.

	The entry offset of \(H\) is odd, since all ordinary block lengths
	are even and all normal offsets are odd. Its initial \(b\) is skipped,
	so the offset in its first \(G\) is one less and hence even. All
	remaining copies of \(G\) in \(H\) are also entered at even offsets.
	They read only \(b\)'s and append only \(b\)'s. The ordinary blocks
	waiting behind \(H\) all have even length, and the padding before
	their internal \(P\) has even length. Each therefore preserves even
	parity and reads only \(b\)'s, both inside \(P\) and in its padding.
	After this finite traversal the dataword consists entirely of \(b\)'s.
	Every further step shortens it by \(\beta-1>0\), so it halts.
	This proves the equivalence and hence the undecidability of halting
	on the designated family.
\end{proof}

\begin{lemma}\label{lem:terminal}
	Every dataword reachable from \(w_0\) is nonempty, ends in \(b\), and
	has length congruent to \(1\) modulo \(\beta-1\). The computation
	halts if and only if it reaches the one-letter word \(b\).
\end{lemma}
\begin{proof}
	The initial length is \((s-1)\beta+1\). A \(b\)-step changes it by
	\(1-\beta\), and a \(c\)-step by \((s-1)\beta\). Both changes preserve
	the congruence. The initial word and both productions end in \(b\),
	so every reachable word is nonempty and ends in \(b\). Its only
	possible length below \(\beta\) is therefore \(1\).
\end{proof}

\section{The PCP(5) instance}\label{sec:pcp-instance}

The PCP(5) simulation uses \(U\) only as a word over
\(\{b,c\}\). It does not decode the cyclic program stored
in that word. Define the morphism
\[
\langle\cdot\rangle:\{b,c\}^*\longrightarrow\{0,1\}^*
\]
by
\[
\langle b\rangle=10^\beta1,\qquad
\langle c\rangle=1,
\]
and extend it by concatenation:
\[
\langle x_1\cdots x_m\rangle
=\langle x_1\rangle\cdots\langle x_m\rangle,
\qquad
\langle\varepsilon\rangle=\varepsilon.
\]
In particular, \(\langle U\rangle\) is obtained by encoding
each letter of \(U\). The PCP pairs reproduce the tag
steps using this encoding, independently of the
interpretation of the individual letters inside \(U\).
The converse shows that every PCP solution yields a halting tag
computation.

A \emph{block of a symbol} is a nonempty word consisting
only of copies of that symbol. A block occurring in a word
is \emph{maximal} if it cannot be extended to either side
by another occurrence of the same symbol.

For the closing pair \(F\), we use Rote's longer block of \(1\)'s
\cite[Section~2.8, footnote~3]{Rote2025}. Every block of \(c\)'s in either
production word has length at most \(|Ub|-1\), since the
production words are \(b\) and \(Ub\), both ending in \(b\).
The same bound holds in every reachable dataword:
the initial word is a suffix of \(Ub\), deleting a prefix
cannot lengthen a block of \(c\)'s, and any remaining
suffix ends in \(b\). Thus no block of \(c\)'s can extend
across the boundary with the appended production word.

Set
\[
B=\langle b\rangle=10^\beta1,\qquad
V=\langle U\rangle,\qquad
L=|Ub|+99.
\]
We form exactly five binary pairs, displayed as
(upper word, lower word):
\begin{equation}
\begin{array}{c|c|c}
\text{pair}&\text{upper}&\text{lower}\\ \hline
A&1&1V10\\
B_+&B&110\\
B_-&B&0\\
C_-&1&0\\
F&10^\beta1^L&1^L.
\end{array}
\label{eq:tiles}
\end{equation}
The first four pairs are Neary's. The longer closing pair
\(F\), suggested by Rote \cite{Rote2025}, replaces his
\((10^\beta1111,1111)\), without increasing the pair count. The proof of Lemma~7 uses only the inequality $L>|Ub|+1$.
To see the issue with the original exponent~$4$, consider a word $w$
beginning with $bccb$. Its overhang $Z(w)$ begins with $10^\beta 1^4$,
so the upper word of the original closing pair matches this prefix. 
The pair \(A\) initializes and later represents a \(c\)-rule;
\(B_+\) represents a \(b\)-rule; \(B_-,C_-\) delete letters.

For \(w=xb\), where \(x\in\{b,c\}^*\), define
\begin{equation}
	Z(w)=\langle x\rangle 10^\beta.
	\label{eq:clean}
\end{equation}
In particular, \(Z(b)=10^\beta\).
The final \(b\) is represented by a truncated \(B\).
Let \(r\) and \(v\) be the upper and lower concatenations
corresponding to a prefix of a PCP solution. If \(v=rZ(w)\),
then the unmatched suffix \(Z(w)\) of the lower concatenation
is called the \emph{lower overhang}. It represents the tag
dataword \(w\).

Starting with \(A\) leaves the lower overhang \(V10\).
For \(j=0,\ldots,\beta-2\), use \(B_-\) if \(u_j=b\),
and \(C_-\) if \(u_j=c\). Their upper words consume
the first \(\beta-1\) encoded letters of \(U\), and their
lower words extend the terminal \(10\) to \(10^\beta\).
The resulting lower overhang is
\[
\langle u_{\beta-1}\cdots u_\ell\rangle 10^\beta
=Z(w_0),
\]
where \(w_0=u_{\beta-1}\cdots u_\ell b\).

At a nonhalting word, use \(A\) or \(B_+\) according
as its first letter is \(c\) or \(b\); then delete the
following \(\beta-1\) letters with \(C_-\) or \(B_-\).
Lemma \ref{lem:step} verifies the exact word identity.
At the halting word \(b\), \(Z(b)=10^\beta\); using
\(F\) makes the two concatenations equal since
\(10^\beta 1^L=10^\beta1^L\). Thus every halting computation yields a PCP solution.
This forward simulation is due to Neary
\cite[proof of Theorem~11]{Neary2015}; the word identities
for each simulated tag step are verified in
Lemma~\ref{lem:step} below. Here we use Rote's longer
closing pair \cite{Rote2025}.

\section{The converse: a PCP solution yields halting}\label{sec:pcp-converse}

In any PCP solution, the upper and lower concatenations
of every prefix are prefixes of the eventual common word.
Thus one is a prefix of the other. When the
current lower concatenation is the upper concatenation
followed by \(z\), and a pair \((a,d)\) has \(a\)
as a prefix of \(z\), the new lower overhang is
\begin{equation} z'=a^{-1}zd. \label{eq:transition}\end{equation}
More generally, if $(a,d)$ is the next pair in a solution, then $a$
and $zd$ must be prefix-comparable, since both are prefixes of the
remaining part of the eventual common word. Hence a disagreement
within their common range cannot be repaired by later pairs.

We first establish the form of any PCP solution
up to its first occurrence of \(F\).

\begin{lemma}\label{lem:blocks}
Every solution starts with \(A\) and ends with \(F\).
Up through its first occurrence of \(F\), it has the form
\begin{equation}
A D_{1,1}\cdots D_{1,\beta-1}\;
T_2D_{2,1}\cdots D_{2,\beta-1}\;\cdots\;
T_kD_{k,1}\cdots D_{k,\beta-1}\;F,
\label{eq:phases}
\end{equation}
where \(T_i\in\{A,B_+\}\) and
\(D_{i,j}\in\{B_-,C_-\}\).
\end{lemma}
\begin{proof}
Only \(A\) can start an equality. Both its words begin
with 1, whereas the lower words of \(B_-,C_-\) start
with 0, and those of \(B_+,F\) start with 11
while their upper words start with 10. Every upper
word ends in 1, whereas only the lower word of
\(F\) ends in 1. Hence every complete match ends in \(F\).

Every maximal block of zeros in an upper word has length
\(\beta\). Since every upper word begins and ends in \(1\),
the same holds for every concatenation of upper words. Every maximal block of zeros in \(V=\langle U\rangle\)
has length \(\beta\).
The lower word of each \(A\) or \(B_+\) ends in
\(10\), starting a zero block of length 1.
Each deletion pair contributes one zero and no
production or closing pair can contribute a zero
at this boundary. Exactly \(\beta-1\) deletion pairs must therefore follow
each occurrence of \(A\) or \(B_+\) before the next
occurrence of \(A\), \(B_+\), or \(F\). Apply this
from the initial \(A\) up to the first \(F\).
A solution may continue after its first occurrence
of \(F\); only the prefix through that occurrence
is considered here.
\end{proof}

The next lemma shows that the initial group of
pairs produces the encoding of the prescribed
tag input.

\begin{lemma}\label{lem:init}
Assume that a PCP solution has the form
\eqref{eq:phases}. After its initial sequence
\[
A D_{1,1}\cdots D_{1,\beta-1},
\]
the lower overhang is \(Z(w_0)\). Moreover, for
\(j=0,\ldots,\beta-2\), the pair \(D_{1,j+1}\)
is uniquely determined by \(u_j\):
\[
D_{1,j+1}=
\begin{cases}
	B_-, & u_j=b,\\
	C_-, & u_j=c.
\end{cases}
\]\end{lemma}
\begin{proof}
	Immediately after \(A\), the lower overhang is
	\(\langle U\rangle 10\), where \(U=u_0\cdots u_\ell\).
	By \eqref{eq:phases}, the next \(\beta-1\) pairs are
	\(D_{1,1},\ldots,D_{1,\beta-1}\), each belonging to
	\(\{B_-,C_-\}\).
	
	We show successively that \(D_{1,j+1}\) is determined
	by \(u_j\), for \(j=0,\ldots,\beta-2\).
	After the first \(j\) deletion pairs have been applied
	as prescribed, the lower overhang is
	\[
	\langle u_j\cdots u_\ell\rangle 10^{j+1}.
	\]
	If \(u_j=b\), the upper word of \(B_-\) consumes
	\(\langle b\rangle=10^\beta1\). Choosing \(C_-\)
	instead would consume only its first \(1\), leaving
	a leading zero. Since every upper word begins in \(1\),
	this choice is incompatible with a PCP solution.
	
	If \(u_j=c\), the overhang begins in \(11\): the first
	\(1\) encodes \(u_j\), and the second begins the next
	code, or the terminal \(10^{j+1}\). Thus \(B_-\),
	whose upper word begins in \(10\), cannot be used,
	and \(C_-\) is forced.
	
	In either case the chosen pair consumes
	\(\langle u_j\rangle\) and appends one zero to the
	lower concatenation. Consequently, after
	\(A D_{1,1}\cdots D_{1,\beta-1}\), the lower overhang is
	\[
	\langle u_{\beta-1}\cdots u_\ell\rangle 10^\beta
	=Z(w_0),
	\]
	as required.
\end{proof}

The next lemma shows that each subsequent group
\(T_iD_{i,1}\cdots D_{i,\beta-1}\) represents
exactly one tag step.

\begin{lemma}\label{lem:step}
	Assume that a PCP solution has the form \eqref{eq:phases},
	and let \(2\leq i\leq k\). Suppose that immediately before
	the occurrence of \(T_i\), the lower overhang is \(Z(w)\),
	where
	\[
	w=x_0x_1\cdots x_{m-1}
	\]
	is a reachable tag dataword with \(m\geq\beta\).
	Then the pairs in the sequence
	\[
	T_iD_{i,1}\cdots D_{i,\beta-1}
	\]
	are uniquely determined by the first \(\beta\) letters
	of \(w\). Specifically,
	\[
	T_i=
	\begin{cases}
		B_+, & x_0=b,\\
		A,   & x_0=c,
	\end{cases}
	\qquad
	D_{i,j}=
	\begin{cases}
		B_-, & x_j=b,\\
		C_-, & x_j=c,
	\end{cases}
	\quad (1\leq j\leq\beta-1).
	\]
	After this sequence, the lower overhang is \(Z(w')\),
	where
	\[
	w'=x_\beta\cdots x_{m-1}R(x_0)
	\]
	is the tag successor of \(w\). When \(m=\beta\), the
	suffix \(x_\beta\cdots x_{m-1}\) is empty.
\end{lemma}
\begin{proof}
	Since \(w\) is reachable, its last letter is \(b\).
	Thus \(x_{m-1}=b\), and
	\[
	Z(w)=\langle x_0\cdots x_{m-2}\rangle 10^\beta.
	\]
	By \eqref{eq:phases}, \(T_i\) is either \(A\) or \(B_+\).
	
	If \(x_0=c\), the lower overhang begins with \(11\):
	the first \(1\) encodes \(x_0\), and the second begins
	the representation of the next letter. The upper word
	of \(B_+\) begins with \(10\), so \(T_i=A\) is forced.
	If \(x_0=b\), the lower overhang begins with
	\(B=10^\beta1\). Choosing \(A\) would consume only its
	first \(1\), leaving a leading zero. Since every upper
	word begins in \(1\), this would be incompatible with
	a PCP solution. Hence \(T_i=B_+\) is forced.
	
	To treat both cases together, put
	\[
	y=
	\begin{cases}
		U, & x_0=c,\\
		\varepsilon, & x_0=b.
	\end{cases}
	\]
	Then \(R(x_0)=yb\). In either case the upper word of
	\(T_i\) is \(\langle x_0\rangle\), and its lower word
	is \(1\langle y\rangle10\). Immediately after \(T_i\),
	the lower overhang is therefore
	\[
	\begin{aligned}
		&\langle x_1\cdots x_{m-2}\rangle
		10^\beta1\langle y\rangle10\\
		&\qquad=
		\langle x_1\cdots x_{m-1}\rangle
		\langle y\rangle10.
	\end{aligned}
	\]
	Here the first \(1\) of the appended lower word completes
	the truncated representation \(10^\beta\) of the old
	final \(b\) to its full code \(B\).
	
	The deletion pairs \(D_{i,1},\ldots,D_{i,\beta-1}\)
	are now forced by the same argument as in the proof
	of Lemma~\ref{lem:init}: for \(1\leq j\leq\beta-1\),
	we must use \(B_-\) if \(x_j=b\), and \(C_-\) if
	\(x_j=c\). Indeed, each of these symbols has its
	full code in the overhang, including the final
	\(b\) when \(m=\beta\).
	
	Each deletion pair consumes \(\langle x_j\rangle\)
	and appends one zero. Hence, after all \(\beta-1\)
	deletion pairs, the lower overhang is
	\[
	\langle x_\beta\cdots x_{m-1}\rangle
	\langle y\rangle10^\beta
	=
	Z(x_\beta\cdots x_{m-1}yb)
	=
	Z(w').
	\]
	Here \(x_\beta\cdots x_{m-1}\) is understood to be
	empty when \(m=\beta\). Thus the sequence represents
	exactly one tag step in every case.
	
	The same identities also show that, starting with
	\(Z(w)\), the prescribed pairs can always be appended
	to obtain \(Z(w')\), without assuming a complete
	PCP solution.
\end{proof}

In the following lemma we show that $F$ can only appear when the reachable tag word is $b$.

\begin{lemma}\label{lem:closing}
	Assume that a PCP solution has the form
	\eqref{eq:phases}. Suppose that immediately before
	its first \(F\), the lower overhang is \(Z(w)\)
	for a reachable tag word \(w\).
	Then \(w=b\), and this occurrence of \(F\)
	makes the upper and lower concatenations equal.
	Conversely, if the lower overhang is \(Z(b)\),
	applying \(F\) gives equality.
\end{lemma}

\begin{proof}
	Recall that the upper and lower words of \(F\)
	are \(10^\beta1^L\) and \(1^L\), respectively,
	where \(L=|Ub|+99\).
	
	If \(w\) starts with \(c\), then \(Z(w)\) begins
	with \(11\), since \(w\) ends in \(b\).
	This is incompatible with the upper word of \(F\),
	which begins with \(10\).
	
Suppose that \(w\) starts with \(b\) and \(|w|>1\). Since \(w\)
ends in \(b\), write \(w=bc^rbt\) for some \(r\ge0\). The run
bound in Section~\ref{sec:pcp-instance} gives \(r\le |Ub|-1\).
After the initial \(10^\beta\) in \(Z(w)\), there are exactly
\(r+2\) consecutive \(1\)'s: one ending the first \(b\)-code,
\(r\) encoding the intervening \(c\)'s, and one beginning the
next \(b\)-code. The next symbol is \(0\), including when that
second \(b\) is the final letter of \(w\). Since
\[
r+2\le |Ub|+1<L,
\]
this \(0\) already disagrees with the upper word \(10^\beta1^L\)
of \(F\). The disagreement lies within the existing overhang.

	The only remaining possibility is \(w=b\).
	In this case \(Z(w)=10^\beta\), and
	\[
	Z(b)\,1^L=10^\beta1^L,
	\]
	which is exactly the upper word of \(F\).
	Hence applying \(F\) makes the two
	concatenations equal.
\end{proof}

Now we are ready to give a full proof for the result of Neary.  

\begin{theorem}
	PCP(5) is undecidable over the binary alphabet.
\end{theorem}
\begin{proof}
	Consider the restricted family of binary tag systems
	described in Lemma~\ref{lem:neary9}, whose halting problem is
	undecidable.
	Given a system in this family, with deletion number
	\(\beta\), productions
	\[
	b\longmapsto b,
	\qquad
	c\longmapsto Ub,
	\]
	and prescribed initial word \(w_0\), form the five
	pairs in \eqref{eq:tiles}.
	All their words are binary, and they can be computed
	effectively from \(\beta\) and \(U\).
	We prove that the resulting PCP instance has a
	solution if and only if the tag system halts
	from \(w_0\).
	
	Suppose first that the tag computation halts. By
	Lemma~\ref{lem:terminal}, its final word is \(b\).
	Write the computation as
	\[
	w_0\longrightarrow w_1\longrightarrow\cdots
	\longrightarrow w_t=b.
	\]
	As shown in Section~\ref{sec:pcp-instance}, starting with \(A\)
	and appending the \(\beta-1\) deletion pairs prescribed there
	leaves the lower overhang \(Z(w_0)\).
	For each step \(w_j\longrightarrow w_{j+1}\),
	append the production pair determined by the
	first symbol of \(w_j\), followed by the
	\(\beta-1\) deletion pairs determined by its next
	\(\beta-1\) symbols.
	The word identities in the proof of Lemma~\ref{lem:step} show that this
	changes the lower overhang from \(Z(w_j)\)
	to \(Z(w_{j+1})\).
	After all \(t\) steps, the lower overhang is
	therefore \(Z(b)=10^\beta\).
	Appending \(F\) makes the upper and lower
	concatenations equal, since
	\[
	Z(b)\,1^L=10^\beta1^L.
	\]
	Thus the PCP instance has a solution.
	
	Conversely, suppose that the PCP instance has
	a solution. By Lemma~\ref{lem:blocks}, its prefix
	through the first occurrence of \(F\) has the form
	\eqref{eq:phases}. By Lemma~\ref{lem:init}, the
	initial group
	\[
	A D_{1,1}\cdots D_{1,\beta-1}
	\]
	leaves exactly the lower overhang \(Z(w_0)\).
	
	Proceed through the remaining groups in this
	factorization. Whenever the current overhang
	is \(Z(w)\) for a reachable nonhalting word \(w\),
	Lemma~\ref{lem:step} forces the next production
	pair and its \(\beta-1\) deletion pairs, and
	shows that the resulting overhang is \(Z(w')\),
	where \(w'\) is the genuine tag successor of \(w\).
	Consequently, these groups represent successive
	steps of the computation starting from \(w_0\).
	
	If any word reached in this way is \(b\), the
	tag computation has halted. Otherwise, since
	there are only finitely many groups before the
	first \(F\), this induction reaches the prefix
	immediately before that occurrence of \(F\).
	Its lower overhang is \(Z(w)\) for a reachable
	tag word \(w\). Lemma~\ref{lem:closing} then
	forces \(w=b\), so the tag computation halts
	in this case as well.
	
	We have therefore obtained an effective reduction
	of the restricted tag-system halting problem
	to PCP(5) over the binary alphabet.
	Since the former problem is undecidable,
	so is the latter.
\end{proof}

The Lean theorem
\texttt{Pcp5Lean.EndToEnd.source\_halts\_iff\_pcp5}
in the companion formalization \cite{leanPCP5} verifies that the
constructed binary PCP instance has exactly five listed pairs and
admits a nonempty solution if and only if the source machine halts.
Its proof composes the repaired cyclic-tag simulation, the conversion
to binary tag, and the reduction to PCP. The formalization verifies
the no-refiring property during consumption of the old residual queue,
which is the interval used by the reduction.

\section{Conclusion}

Neary's construction establishes the undecidability
of PCP(5) over the binary alphabet. Our purpose
has been to give a detailed verification of this
result, with particular attention to the converse:
every PCP solution yields a halting computation
of the simulated tag system.

Lemmas~\ref{lem:blocks}--\ref{lem:closing}
make this implication explicit. The lengths of
maximal blocks of zeros force the form
\eqref{eq:phases}; comparison with the lower
overhang then forces the initialization and
each successive tag step. Finally, the closing
pair can be applied only when the represented
tag word is \(b\).

We use Rote's longer closing pair
\cite{Rote2025}, with \(L=|Ub|+99\), in place
of the four consecutive \(1\)'s in Neary's
original pair. Its required property follows
from the bound on the lengths of blocks of
\(c\)'s in reachable tag words.

The undecidability of the restricted tag-system
family rests on Neary's construction.
The PCP verification uses the tag-system properties established
in Section~\ref{sec:tag}. The internal structure of \(U\) is used
only in the proof of the restricted halting problem. The construction is due to Neary, with the
modification of the closing pair due to Rote;
the contribution here is a detailed exposition
and verification of the resulting reduction
in both directions.

\bigskip

\paragraph{Acknowledgements.}
I thank Turlough Neary for explaining his construction
to me during a visit to Turku many years ago.
That discussion was an important source of my confidence
in the result and helped motivate the present exposition.

I used OpenAI's ChatGPT to assist with drafting,
revising the exposition, and checking details of the
arguments. I have reviewed the resulting text and
take full responsibility for its mathematical content
and any remaining errors.

The author thanks \href{https://agentformers.fi/}{AgentFormers}
for providing the Lean agents and workstations used to develop
and check the companion formalization \cite{leanPCP5}.

\bibliographystyle{plainurl}
\bibliography{PCP5_references}

@article{Claus,
  author  = {Claus, V.},
  title   = {Some remarks on {PCP}(k) and related problems},
  journal = {Bullet. EATCS},
  volume  = {12},
  pages   = {54--61},
  year    = {1980},
  url     = {https://page.mi.fu-berlin.de/rote/Kram/Volker_Claus-Some_remarks_on_PCP(k)_and_related_problems-1980-Bull-EATCS-12.pdf}
}

@article{MS2005,
  author  = {Matiyasevich, Y. and S{\'e}nizergues, G.},
  title   = {Decision problems for semi-{Thue} systems with a few rules},
  journal = {Theoret. Comput. Sci.},
  volume  = {330},
  number  = {1},
  pages   = {145--169},
  year    = {2005},
  doi     = {10.1016/j.tcs.2004.09.016},
  url     = {https://doi.org/10.1016/j.tcs.2004.09.016}
}

@inproceedings{Neary2015, 
  author    = {Neary, T.},
  title     = {Undecidability in binary tag systems and the {Post} correspondence problem for five pairs of words},
  booktitle = {{STACS} 2015},
  series    = {{LIPIcs}},
  volume    = {30},
  pages     = {649--661},
  year      = {2015},
  doi       = {10.4230/LIPIcs.STACS.2015.649},
  url       = {https://doi.org/10.4230/LIPIcs.STACS.2015.649}
}

@InProceedings{NearyWoods2006,
	author="Neary, T.
	and Woods, D.",
	title="P-completeness of Cellular Automaton Rule 110",
	booktitle="{ICALP} 2006",
 series    = {{LNCS}},
 volume    = {4051},
	year="2006",
	publisher="Springer Berlin Heidelberg",
	pages="132--143",
  doi       = {10.1007/11786986_13},
url       = {https://doi.org/10.1007/11786986_13}
}

@article{PoSt1946,
  author  = {Post, E. L.},
  title   = {A variant of a recursively unsolvable problem},
journal = {Bull. Amer. Math. Soc.},  
volume  = {52},
  pages   = {264--268},
  year    = {1946},
  doi     = {10.1090/S0002-9904-1946-08555-9},
  url     = {https://doi.org/10.1090/S0002-9904-1946-08555-9}
}

@misc{leanPCP5,
	author       = {Halava, V.},
	title        = {{Lean} Formalization of {PCP}(5) Undecidability},
	year         = {2026},
	howpublished = {Zenodo},
	note         = {Version 1.0.0},
	doi          = {10.5281/zenodo.24778217},
	url          = {https://doi.org/10.5281/zenodo.24778217}
}

@inproceedings{Rote2025,
  author    = {Rote, G.},
  title     = {Probabilistic finite automaton emptiness is undecidable for a fixed automaton},
  booktitle = {{MFCS} 2025},
  series    = {LIPIcs},
  volume    = {345},
  pages     = {86:1--86:18},
  year      = {2025},
  doi       = {10.4230/LIPIcs.MFCS.2025.86},
  url       = {https://doi.org/10.4230/LIPIcs.MFCS.2025.86},
}
\end{document}